\documentclass[journal]{IEEEtran}

\AtBeginDocument{%
   \setlength\abovedisplayskip{5pt}
   \setlength\belowdisplayskip{5pt}}

\usepackage{graphicx}
\usepackage{subfig}
\usepackage{amsmath}
\usepackage{amssymb}
\newcommand{\R}{\mathbb{R}}
\usepackage{bm}

\usepackage{float}
\usepackage{comment}
\usepackage{url}
\usepackage[utf8]{inputenc}
\DeclareUnicodeCharacter{2212}{-}
\usepackage{array,multirow}
\usepackage{algorithm,algcompatible,amsmath}
\algnewcommand\INPUT{\item[\textbf{Input:}]}%
\algnewcommand\PARAMS{\item[\textbf{Parameters:}]}
\algnewcommand\STEPONE{\item[{Step 1:}]}
\algnewcommand\STEPTWO{\item[{Step 2:}]}
\algnewcommand\STEPTHREE{\item[{Step 3:}]}
\usepackage{tikz}

\usepackage{caption}
\usepackage{tabularx}
\usepackage{booktabs}
\newcommand{\ra}[1]{\renewcommand{\arraystretch}{#1}}
\usepackage{tikz}
\usepackage{pgfplots}
\pgfplotsset{width=10cm,compat=1.15}
\usepgfplotslibrary{statistics}
\usepgfplotslibrary{colormaps}
\newcommand{\bbox}{\protect\raisebox{1pt}{\protect\tikz \protect\draw[black,fill=black] (1,1) circle (0.5ex);}}
\newcommand{\wbox}{\protect\raisebox{1pt}{\protect\tikz \protect\draw[black,fill=white] (1,1) circle (0.5ex);}}
\usepackage{cite}
\usepackage{dblfloatfix}
\begin{document}

\title{Cross-Modal Knowledge Transfer via Inter-Modal Translation and Alignment for Affect Recognition}

\author{Vandana Rajan,
        Alessio Brutti,
        and Andrea Cavallaro
\thanks{V.~Rajan and A.~Cavallaro are with the Centre for Intelligent Sensing, Queen Mary University of London, UK (e-mail: v.rajan@qmul.ac.uk, a.cavallaro@qmul.ac.uk).}
\thanks{A.~Brutti is with the Fondazione Bruno Kessler, Italy (e-mail: brutti@fbk.eu).}}

\markboth{Journal of \LaTeX\ Class Files,~Vol.~14, No.~8, August~2015}%
{Shell \MakeLowercase{\textit{et al.}}: Bare Demo of IEEEtran.cls for IEEE Journals}

\maketitle

\begin{abstract}
Multi-modal affect recognition models leverage complementary information in different modalities to outperform their uni-modal counterparts. However, due to the unavailability of modality-specific sensors or data, multi-modal models may not be always employable. For this reason, we aim to improve the performance of uni-modal affect recognition models by transferring knowledge from a better-performing (or stronger) modality to a weaker modality during training. Our proposed multi-modal training framework for cross-modal knowledge transfer relies on two main steps. First, an encoder-classifier model creates task-specific representations for the stronger modality. Then, cross-modal translation generates multi-modal intermediate representations, which are also aligned in the latent space with the stronger modality representations. To exploit the contextual information in temporal sequential affect data, we use Bi-GRU and transformer encoder. We validate our approach on two multi-modal affect datasets, namely CMU-MOSI for binary sentiment classification and RECOLA for dimensional emotion regression. The results show that the proposed approach consistently improves the uni-modal test-time performance of the weaker modalities.
\end{abstract}

\begin{IEEEkeywords}
multi-modal-training uni-modal-testing, cross-modal knowledge transfer, emotion recognition, sentiment classification
\end{IEEEkeywords}

\IEEEpeerreviewmaketitle

\section{introduction} \label{sec-1}

\IEEEPARstart{H}{umans} express their affective state through speech, facial expressions and body gestures. 
 However, the information to fully infer the underlying emotional state is often unevenly spread across these modalities. Consequently, acoustic (speech), visual (facial expressions) and textual (linguistic features) modalities exhibit considerable  performance variations for emotion recognition~\cite{han2019emobed} and sentiment classification~\cite{seo2020hmtl}. For example, the level of emotional activation (arousal) is more evident from speech than from facial expressions, whereas the level of emotional positiveness (valence) can be better inferred  from facial expressions than from speech~\cite{valstar2016avec, han2019emobed,  triantafyllopoulos2018audeering, kollias2020exploiting, yin2020speaker}. Similarly, linguistic features extracted from text transcripts can be more informative about the expressed sentiment compared to acoustic or visual features~\cite{poria2017context,seo2020hmtl}.

Multi-modal affect recognition models aggregate the unevenly distributed, complementary information across the available modalities to outperform  uni-modal models~\cite{MorencyTutorial,caridakis2007multimodal,ngiam2011multimodal,tzirakis2017end,hu2019dense}. However, several applications or use cases need uni-modal models. For example, emotion recognition models for emergency service lines~\cite{lefter2011automatic} or customer-support call centers~\cite{segura2016automatic} can rely on speech data only. Furthermore,   drowsiness detection models~\cite{jabbar2018real,reddy2017real} use only face images to detect the level of lethargic state of a driver. These uni-modal models are limited by the characteristics of their respective modality and their performance often falls short of their multi-modal counterparts.  

In this paper, we propose a cross-modal knowledge transfer framework that exploits the complementary information provided by multiple modalities during training in order to improve the test-time performance of uni-modal models~\cite{han2019emobed,seo2020hmtl}. Given a task and two modalities, we define the better performing modality  as {\em stronger} and the other as  {\em weaker}~\cite{zhao2020knowledge, rajan2021robust}. Furthermore, we use the term  {knowledge} to refer to task-specific information available in the features of a modality. 

\begin{table*}[!t]
\centering\ra{1.0}
\setlength{\tabcolsep}{5pt}
\caption{Summary of cross-modal knowledge transfer methods used in multi-modal training and uni-modal testing models. KEY -   $\mathcal{I}$: RGB image; $\mathcal{D}$: depth image; $\mathcal{F}$: optical flow image; $\mathcal{V}$: RGB video; enc.: encoder, dec.: decoder, attn.: attention, trans.: transformer, MLP: multi-layer perceptron, T-S: teacher-student}
\begin{tabular}{@{}clcccccclcl@{}}
\toprule
Reference & Architecture & \multicolumn{6}{c}{Modalities} & \multicolumn{1}{c}{Method} & Sequential & \multicolumn{1}{c}{Task} \\
\cmidrule{3-8}
& & \multicolumn{4}{c}{Visual} & Acoustic & Textual & & & \\
\cmidrule{3-6}
&  & $\mathcal{I}$ & $\mathcal{D}$  & $\mathcal{F}$  & $\mathcal{V}$  & & & & & \\
\midrule
\cite{abavisani2019improving} & I3D (3D CNN) & \bbox & \bbox & \bbox & \wbox & \wbox & \wbox & correlation alignment & \bbox & gesture recognition \\
\cite{garcia2019learning} & Resnet-50 & \bbox & \bbox & \wbox & \wbox & \wbox & \wbox &  weight initialisation, & \bbox & action recognition, \\
& & & & & & & & adversarial training & & object classification \\
\cite{aguilar2019multimodal} & LSTM-attn. & \wbox & \wbox & \wbox & \bbox & \bbox & \bbox & contrastive loss & \bbox & emotion recognition \\
\cite{albanie2018emotion} & CNN & \wbox & \wbox & \wbox & \bbox & \bbox & \wbox & T-S knowledge distillation & \bbox & emotion recognition \\
\cite{pham2018seq2seq2sentiment} & LSTM enc.-dec. & \wbox & \wbox & \wbox & \bbox & \bbox & \bbox & cross-modal translation & \bbox & sentiment analysis \\
\cite{han2019emobed} & GRU & \wbox & \wbox & \wbox & \bbox & \bbox & \wbox & joint audio-visual training, & \wbox & emotion recognition \\
& & & & & & & & cross-modal triplet training & & \\
\cite{seo2020hmtl} & GRU-attn. & \wbox & \wbox & \wbox & \bbox & \bbox & \bbox & cross-modal translation, & \bbox & sentiment analysis, \\
& & & & & & & & adversarial training & & emotion recognition \\
\cite{dumpala2019audio} & MLP & \wbox & \wbox & \wbox & \bbox & \bbox & \wbox & cross-modal translation, & \wbox & sentiment analysis \\
& & & & & & & & correlation alignment & & \\
\cite{rajan2021robust} & MLP & \wbox & \wbox & \wbox & \bbox & \bbox & \wbox & cross-modal translation, & \wbox & emotion recognition \\
& & & & & & & & correlation alignment & & \\
Ours & GRU-trans.enc. & \wbox & \wbox & \wbox & \bbox & \bbox & \bbox & cross-modal translation, & \bbox & sentiment analysis,  \\
& & & & & & & & correlation alignment & & emotion recognition\\
\bottomrule
\end{tabular}
\label{tab:summary-sota}
\end{table*}
The proposed framework,  \textit{Cross-Modal Stronger Enhancing Weaker} (\mbox{CM-StEW}), is a supervised neural network training strategy that improves the test-time performance of a weaker modality by exploiting a stronger modality  \textit{during training phase only}. 
A pre-trained uni-modal stronger modality model transfers its knowledge to a uni-modal weaker modality model, helping the latter learn better discriminative representations. 
Cross-modal knowledge transfer is achieved by combining weaker-to-stronger modality translation and feature alignment. The main idea is that cross-modal translation can create intermediate representations that capture joint information between the modalities~\cite{pham2018seq2seq2sentiment,wang2020transmodality}. In addition to this, an explicit alignment between the intermediate and  the stronger modality representations further encourages the framework to discover components of the weaker modality that are maximally correlated with the stronger modality. Unlike our previous work SEW~\cite{rajan2021robust}, CM-StEW uses a supervised encoder-classifier model for the stronger modality that can create better task-specific representations and can deal with sequential data. 
The transformer encoder based architecture and the Bi-GRU front-end of CM-StEW lead to new state-of-the-art results for the acoustic and visual uni-modal sentiment analysis on the CMU-MOSI dataset, as well as for continuous emotion recognition on the RECOLA dataset.

\section{Related work} 
\label{sec-2}

In this section, we discuss cross-modal knowledge transfer frameworks applied to different multi-modal tasks, including affect recognition (see Tab.~\ref{tab:summary-sota}). We also discuss related works on multi-modal sentiment classification and continuous emotion recognition.

\subsection{Cross-modal knowledge transfer}
Several cross-modal knowledge transfer frameworks for enhancing uni-modal performance were developed for applications where the  modalities are different types of images (e.g.~RGB and depth). For example, a multi-modal training framework, called MTUT, uses RGB, depth and/or optical flow images to improve uni-modal hand gesture recognition models that use either of the three modalities at test-time~\cite{abavisani2019improving}. The modality specific parts of the model use deep 3D CNN architecture~\cite{carreira2017quo}, and are forced to derive common semantics from the spatio-temporal contents of RGB, depth and/or optical flow images. Another framework, called ADMD, uses knowledge distilled from depth images to improve the uni-modal test-time performance of RGB images for action recognition and object classification~\cite{garcia2019learning}. Weights of the RGB stream are initialised using a pre-trained depth stream and are trained using an adversarial loss feature alignment between RGB and depth representations. Both modality specific streams use the Resnet-50~\cite{he2016deep} architecture. The modalities considered in these works capture the same visual scene and the uni-modal models share the same architecture, which help in creating a common semantic understanding of the scene between the modalities. Hence, these solutions are not directly applicable for knowledge transfer between modalities  that capture different spatio-temporal semantics even when they correspond to the same physical event (e.g.~acoustic, visual, textual). 

Works for cross-modal knowledge transfer using acoustic, visual and/or textual modalities exploit the complementary information from these modalities during training to develop a uni-modal ~\cite{han2019emobed,pham2018seq2seq2sentiment,albanie2018emotion,aguilar2019multimodal,dumpala2019audio} or bi-modal~\cite{seo2020hmtl} system. A joint audio-visual training and cross-modal triplet loss based framework, called EmoBed, can be used for multi-modal training in face/speech emotion recognition tasks~\cite{han2019emobed}. An inherent caveat in such a framework can be the degradation in performance of the stronger modality caused by the weaker modality as reported in~\cite{han2019emobed}. A recent work, HMTL, uses a cross-modal decoder and discriminator to transfer knowledge from text modality to audio/visual modality for sentiment classification~\cite{seo2020hmtl}. However, the use of discriminator based adversarial training demands additional parameters along with added complexities such as oscillations in loss values during training~\cite{goodfellow2016nips}. An unsupervised cross-modal translation based framework, called Seq2SeqSentiment, uses an LSTM based encoder-decoder model to create intermediate features that are representative of both modalities~\cite{pham2018seq2seq2sentiment}. However, the absence of supervision during translation can create representations that might not be discriminative for the task at hand. Knowledge from a bi-modal audio-text model can be induced into an audio-only emotion recognition model using contrastive loss \cite{aguilar2019multimodal}. This method is specifically designed for knowledge transfer from text-audio model to audio-only model as it uses words to derive semantic audio features. An unsupervised knowledge transfer framework, called DCC-CAE, uses correlation based feature alignment with cross-modal translation to develop uni-modal audio or visual based sentiment classification model \cite{dumpala2019audio}. DCC-CAE neither takes contextual information into account nor utilises the superior knowledge from text modality. Also, unsupervised feature alignment can result in representations which are sub-optimal for the classification task. Another type of cross-modal knowledge transfer framework for speech emotion recognition uses a strong teacher model trained on labelled visual modality to distill knowledge to the student model trained on unlabelled speech modality~\cite{albanie2018emotion}. However, this method requires a teacher model, that has to be pre-trained using large amount of labelled face images, possibly using multiple datasets. Also, the results indicate that the student model's performance falls short of speech emotion recognition models which were trained in a fully supervised manner. 

\subsection{Affect recognition}
Discrete sentiment classification and continuous emotion regression are two common affect recognition tasks in literature that use one or more modalities to infer the underlying sentiment or emotion expressed by a human subject~\cite{poria2017review}. While sentiment analysis task focuses on classification of the polarity of opinion or feeling (positive or negative) expressed by the human subject, continuous emotion regression task aims to obtain finer interpretations of the underlying emotional state of the person (like the level of activation or arousal and the level of positiveness or valence). 

Over the years, researchers have applied various creative techniques to effectively model sentiment information using acoustic, visual and textual modalities. 1D CNN and RNN (uni or bi-directional LSTM or GRU) layers are commonly used to exploit local and global temporal information respectively~\cite{tsai2019multimodal,poria2017multi,ghosal2018contextual,seo2020hmtl}. An attention operation can be applied over sequential features of the same modality to capture intra-modality dynamics by providing varying focus to different temporal segments in the same sequence~\cite{poria2017multi,ghosal2018contextual,seo2020hmtl}. Inter-modality relationships can also be exploited for multi-modal fusion by using cross-modal attention~\cite{ghosal2018contextual} or transformer encoders~\cite{tsai2019multimodal}. However, analysis of the attention maps as well as ablation studies on the multi-modal models provide evidence that they rely more on the superior textual modality than the weaker acoustic or visual modalities. 

Continuous emotion recognition task has been popularised by the AVEC challenges over the years since 2016~\cite{valstar2016avec,ringeval2018avec}. AVEC 2016 provides a baseline model using Support Vector Regressor (SVR). Since then, researchers have introduced various deep learning based uni-modal and multi-modal models. An RNN based auto-encoder can be used to reconstruct input features and the reconstruction error (RE) thus obtained along with the original features can be input to another RNN model to obtain final emotion predictions~\cite{han2017reconstruction}. Similar to RE, the perceptual uncertainty (PU) associated with label annotation can also be used to obtain more accurate emotion predictions~\cite{han2017hard}. A dynamic difficulty awareness training framework (DDAT) uses either RE or PU to obtain an estimate of how much difficulty is associated with the prediction of each data sample. This difficulty estimate can then be used along with the input features to train a continuous emotion recognition model~\cite{zhang2018dynamic}. Even though these models utilise multi-modal fusion techniques to improve their final emotion prediction performance, during deployment phase, they demand the simultaneous presence of all the modalities that were present during the training stage. 

In contrast to the above methods, we aim to solve a different problem; how to improve the uni-modal test-time performance of a weaker modality by transferring knowledge from a superior or stronger modality during training phase. 

\section{cross-modal stronger enhancing weaker} 
\label{sec-4}

\subsection{Modality ranking} \label{mod-rank}
Let a human subject displaying affect be captured in a multi-modal clip \mbox{$ V = [V_1,V_2,.....,V_N] $} that is split into $N$ segments. Each segment $V_n\;(n=\{1, \dots, N\})$ is defined by a fixed temporal window or by utterance boundaries. Let the corresponding labels be represented by \mbox{$Y = [Y_1, Y_2, ...., Y_N] \in \R^{N \times 1}$}.

Let $a$, $v$ and $t$ represent acoustic, visual and textual modalities, respectively. The sequence of features for modality $m \in \{a,v,t\}$ are given by \mbox{$X_m = [X_{m_1}, X_{m_2}, ...., X_{m_N}] \in \R^{N \times d_m}$}, where $d_m$ is the feature dimensionality. 

Let $\Gamma\rule{0pt}{7.45pt}^{c}$ ($\Gamma\rule{0pt}{7.45pt}^{r}$) denote the uni-modal classifier (regressor) with parameters $\theta_m$ for modality $m$. Let the performance 
score for each modality, $e_m$, be given by the evaluation metric $\mathcal{E}$ as
\begin{equation}
e_m = \mathcal{E}\left(\Gamma\rule{0pt}{7.45pt}^{c}(X_m;\theta_{m}),Y\right).
\end{equation}

Using this performance score, we rank  the modalities for a specific classification or regression task:  a modality $s$ is said to be stronger than modality $w$ if $e_s > e_w$ (the opposite if $\mathcal{E}$ measures errors). Then our objective is to improve the task performance of feature $X_w$ (the weaker modality) using the stronger modality during training. The resulting, enhanced feature representation, $X'_w$, will improve the test-time performance compared with $X_w$.

\subsection{Stronger modality encoder-classifier training} \label{step-1}

We first create a model for the stronger modality, which acts as the `source' of knowledge to be transferred to another modality. This source model consists of an encoder and a classifier. The purpose of encoder is to effectively model the affect information contained in the input stronger modality features and to map them into a latent space of desired dimensionality. The classifier ensures that the latent representations thus obtained are discriminative for the specific task. 

The affect information of a given segment $V_n$ often has inter-dependence on other neighbouring contextual segments in the same clip $V$~\cite{ghosal2018contextual,poria2017context}. To model this inter-dependence across time, we use Bi-GRU as the first layer of our encoder to transform the input feature sequence into `context aware' representations, which are then input to a dense layer. The dense (fully connected) operation is shared across the time-steps, for projecting `context aware' features onto a fixed dimension $d$. Even though the Bi-GRU layers can capture contextual information via the hidden states, they cannot provide varying focus on which hidden states carry more valuable affect information. This can be accomplished by the use of self-attention mechanism that provides varying levels of attention weights to the time-steps in the same sequence. For this purpose, we exploit the stacked self-attention mechanism using transformer encoder layers~\cite{vaswani2017attention}. The use of Multi Head Attention (MHA), which contains multiple self-attention operations, allows the transformer to capture richer interpretations of the input sequence. 

\begin{figure}[t!]
\centering
\includegraphics[width=9.1cm,height=12cm]{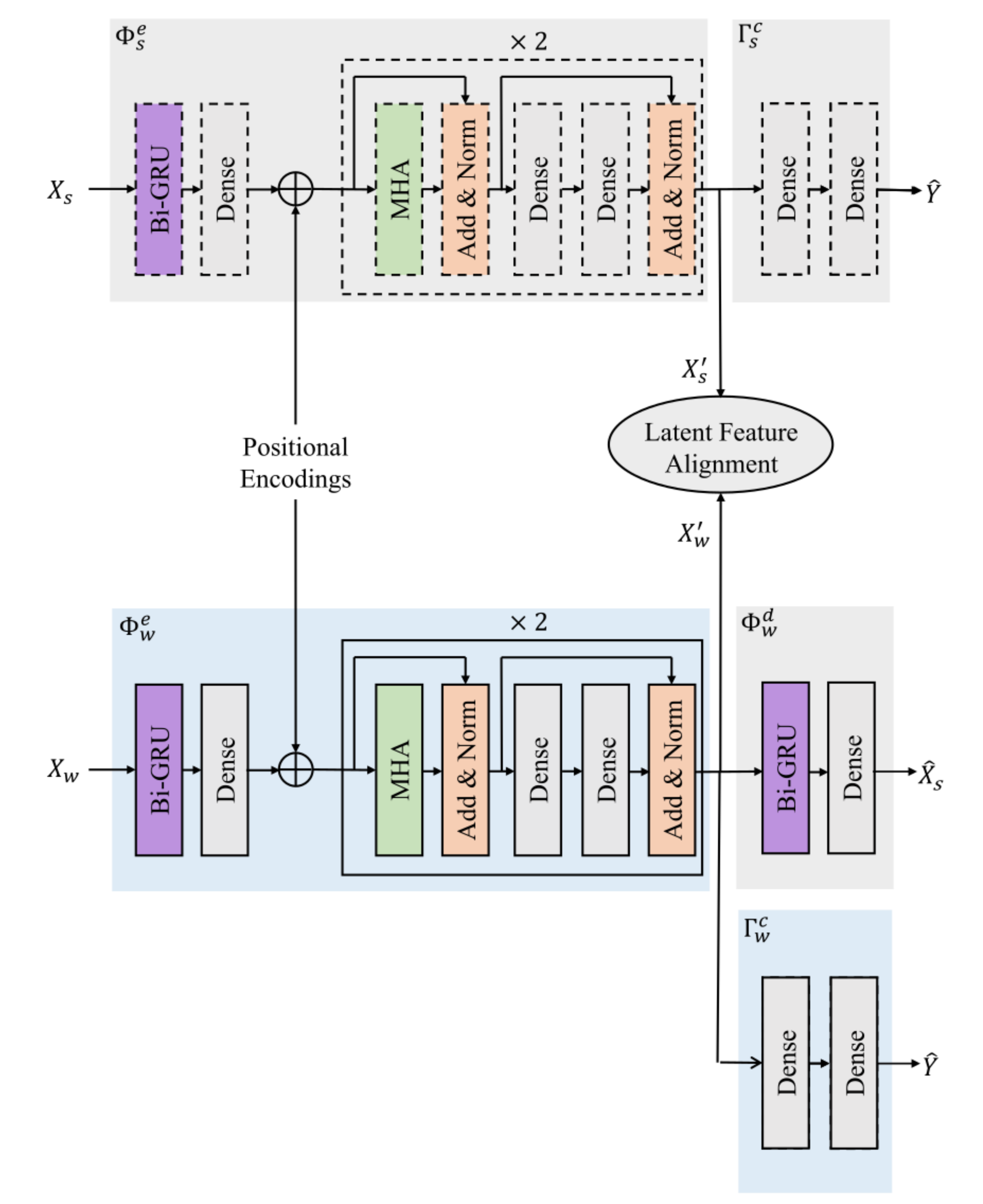}%
\caption{CM-StEW model. $X_s$ and $X_w$ are stronger and weaker modality inputs, $\hat{Y}$ denotes predicted labels, $X'_s$ and $X'_w$ are the two intermediate representations, $\hat{X}_{s}$ denotes the reconstructed stronger modality features. $\times 2$ shows that two transformer encoder layers are used. Dashed lines indicate layers whose parameters are fixed. Only the blocks in blue background ($\Phi_w^e$ and $\Gamma_w^c$) have to be retained after the training. KEY: MHA-multi-head attention, Add: addition, Norm: layer normalisation.}
\label{fig:Model}
\end{figure}
%
%The MHA module in each transformer encoder layer is followed by two dense layers with a ReLU activation in between them. These dense layers are shared across the time-steps and their purpose is to process the output from one MHA  module in a way to better fit the input for the next MHA module of a multi-layered transformer encoder. Furthermore, residual addition and layer normalisation is done at the output of both the MHA module as well as the dense layers to ease the gradient flow through the multi-layered transformer encoder and to obtain faster training and better generalization ability. 
Note that the transformer maintains same feature dimensionality ($d$) at the output of both MHA module as well as the dense layers to facilitate residual addition. Additionally, for the transformer encoder layers to be aware of the temporal order of the input sequence, positional information in the form of sinusoidal position embeddings is added to the input of the first transformer encoder layer~\cite{vaswani2017attention}.

As output of the multi-layer transformer encoder, we obtain attention-weighted and context aware stronger modality representations of a desired dimensionality. In order to ensure that these representations are discriminative for the specific task, they are fed into a classifier made up of two dense layers. The first dense layer has Relu activation whereas the second layer has Sigmoid or Tanh activation for classification or regression task respectively. 

The upper part of Fig.\ref{fig:Model} shows the encoder and classifier of source model as $\Phi_s^e$ and $\Gamma^c_s$. $\Phi_s^e$ takes a sequence of stronger modality features \mbox{$X_s \in \R^{N \times d_s}$} as input and converts them into attention-weighted, context aware latent representations \mbox{$X'_s \in \R^{N \times d}$}. Note that the encoder has brought the feature dimension from $d_s$ to $d$. The classifier $\Gamma^c_s$ then takes  $X'_s$ as input and provides the predicted labels $\hat{Y} \in \R^{N \times 1}$. 

We train the entire source model in an end-to-end fashion to map the input stronger modality features into the task-specific label space. Following~\cite{han2019emobed, seo2020hmtl}, we use Mean-Square-Error or Binary-Cross-Entropy for training depending upon whether the task is regression or binary classification respectively. If $\hat{Y}_n$ denotes the predicted label for segment $n$ at the classifier output and $Y_n$ represents the corresponding true label, then the prediction loss for regression and classification are given by,
\begin{equation} \label{lp:mse}
    \mathcal{L}_p = \frac{1}{N}\sum_{n=1}^{N}(Y_n-\hat{Y}_n)^2
\end{equation}
and
\begin{equation} \label{lp:bce}
    \mathcal{L}_p = \frac{1}{N}\sum_{n=1}^{N}(Y_n.\log(\hat{Y}_n)+(1-Y_n).\log(1-\hat{Y}_n)) % BCE
   % \mathcal{L}_p = \frac{1}{N}\sum_{n=1}^{N}(Y_n . \log(\hat{Y}_n))
\end{equation}
respectively, where $N$ represents the total number of segments in the sequence. Once the training is over, the source model parameters are fixed and only the encoded representations $X'_s$ are retrieved for the next steps of knowledge transfer.

\subsection{Cross-modal translation and feature alignment}
CM-StEW uses a combination of cross-modal translation and correlation based latent feature alignment for knowledge transfer from the pre-trained stronger modality source model to the weaker modality model. Cross-modal translation from the weaker to stronger modality creates intermediate features that are representative of both the modalities. A correlation based latent feature alignment increases the correlation between these intermediate representations and the stronger modality representations obtained from the source model. This helps the weaker modality model discover components of weaker modality features that are more discriminative for the task. 

Similar to the source model, the weaker modality model also uses an encoder and a classifier to obtain attention-weighted, context aware and task-specific discriminative features. The encoder maps the weaker modality features into latent representations of same dimensionality $d$ as the encoded representations from the source model. The encoder and classifier architecture are the same as the source model. Additionally, a decoder is used to map the output of encoder to the stronger modality features. Thus, the encoder output is given as input to both the decoder as well as the classifier. To facilitate the sequential aspect of cross-modal translation, the decoder is made up of Bi-GRU layers, which are followed by a dense layer. The dense operation is shared across the time-steps, for projecting the representations from each time step onto the same dimension as the stronger modality features.

For latent feature alignment, we use Deep Canonical Correlation Analysis (DCCA)~\cite{andrew2013deep}, to increase the correlation between intermediate features ($X'_w$) and stronger modality representations obtained from the source model ($X'_s$). If ${\overline{X}'_s}$ and ${\overline{X}'_w}$ are the mean-centred versions of $X'_s$ and $X'_w$ respectively, then the total correlation of $X'_s$ and $X'_w$ is the sum of all singular values (or trace norm) of the matrix, \mbox{\begin{math} T = \Sigma_{s}^{-1/2} \Sigma_{sw} \Sigma_{w}^{-1/2} \end{math}} and is given by,
\begin{equation}
\begin{split}
    corr(X'_s, X'_w) & = ||T||_{tr} \\
                     & = tr(T^\intercal T)^{\frac{1}{2}}.
\end{split}
\end{equation}
The self (\begin{math} \Sigma_{s}, \Sigma_{w} \end{math}) and cross covariance (\begin{math} \Sigma_{sw} \end{math}) matrices are given by
\begin{equation} \label{eq:7}
    \Sigma_{sw} = \frac{1}{N-1} {\overline{X}'_s}^\intercal \ {\overline{X}'_w},
\end{equation}
\begin{equation} \label{eq:8}
    \Sigma_{s} = \frac{1}{N-1} {\overline{X}'_s}^\intercal \ {\overline{X}'_s} + r_1 I,
\end{equation}
\begin{equation} \label{eq:9}
    \Sigma_{w} = \frac{1}{N-1} {\overline{X}'_w}^\intercal \ {\overline{X}'_w} + r_2 I,
\end{equation}
where $r_{1}$ $>$ 0 and $r_{2}$ $>$ 0 are the regularisation constants, $I$ is the identity matrix of size $d \times d$.

The lower part of Fig.~\ref{fig:Model} shows the encoder, decoder and classifier as $\Phi^e_w$, $\Phi^d_w$ and $\Gamma^c_w$ respectively. $\Phi^e_w$ takes a sequence of weaker modality features $X_w \in \R^{N \times d_w}$ as input and converts them into attention-weighted, context aware representations $X'_w \in \R^{N \times d}$. The encoder brings the dimensionality of the features from $d_w$ to $d$ for compatibility with the source model encoder output. $\Phi^d_w$ takes these intermediate representations $X'_w$ as input and translates them into the stronger modality features $X_s$. The back-end $\Gamma^c_w$ maps the enhanced features $X'_w$ into the label space, solving the regression or classification task. 

The encoder, decoder and classifier are jointly trained by optimising three objective functions: a translation loss between the decoder output and the stronger modality features, alignment loss between the encoder output and the stronger modality representations obtained from the source model and a task-specific prediction loss between the true and predicted labels. Following~\cite{pham2018seq2seq2sentiment}, we use Mean-Absolute-Error as cross-modal translation loss. If $X_{s_n}$ and $\hat{X}_{s_n}$ denote the stronger modality features and the decoder output for segment $n$ respectively, then, translation loss $\mathcal{L}_t$ is given by,
\begin{equation}
    \mathcal{L}_t = \frac{1}{N}\sum_{n=1}^{N}|X_{s_n}-\hat{X}_{s_n}|.
\end{equation}
For alignment loss $\mathcal{L}_a$, we use the negative of correlation obtained using DCCA. Thus, 
\begin{equation}
    \mathcal{L}_a = -corr(X'_s, X'_w).
\end{equation}

Similar to the source model, we use Mean-Square-Error or Binary-Cross-Entropy as prediction loss $\mathcal{L}_p$ depending upon whether the task is regression or classification respectively. Thus, $\mathcal{L}_p$ is same as in  eq. \ref{lp:mse} and eq. \ref{lp:bce}. Hence the total training loss is given by,
\begin{equation}\label{eq:total}
    \mathcal{L} = \mathcal{L}_{p} + \alpha\mathcal{L}_{a} + \beta\mathcal{L}_{t},
\end{equation}
where $\alpha$, $\beta$ are scalar weighting hyper-parameters. Once the training is over, only the encoder and classifier of the weaker modality model are to be retained for the inference or deployment phase.

%%%%%%%%%%%%%%%%%%%%%%%%%%%%%%
\section{Experimental Analysis}
\label{sec-5}

We evaluate CM-StEW on two tasks, namely  binary sentiment classification on CMU-MOSI and dimensional emotion regression on RECOLA. We compare CM-StEW with four uni-modal models and two cross-modal knowledge transfer frameworks for binary sentiment classification~\cite{ghosal2018contextual,pham2018seq2seq2sentiment,poria2017multi,seo2020hmtl} and five uni-modal models and two cross-modal knowledge transfer frameworks for emotion regression~\cite{valstar2016avec, han2017reconstruction, han2017hard, zhang2018dynamic, han2019emobed, rajan2021robust}. We also perform an ablation analysis to evaluate the individual contributions of cross modal translation and latent feature alignment in CM-StEW.

For the sentiment classification task, we consider binary accuracy (\textit{Acc.}) and weighted $F_1$ ($\overline{F}_{1}$) scores as performance measures. Binary classification accuracy is the ratio of the correct predictions to the total number of input samples. Weighted $F_1$ is average of $F_1$ (the harmonic mean of precision and recall) for each class weighted by the number of true instances for each class, and is defined as:
\begin{equation}
    \overline{F}_{1} = \frac{F_{1p}n_p + F_{1n}n_n}{n_p+n_n}, 
\end{equation}
where $F_{1p}$ and $F_{1n}$ represent $F_1$ scores for positive and negative classes respectively and $n_p$ and $n_n$ denote the actual number of positive and negative samples respectively.

For the continuous emotion regression task, we use Concordance Correlation Coefficient (CCC), which is defined as:
\begin{equation} \label{eq:ccc}
    \mbox{CCC} = \frac{2\sigma_{Y\hat{Y}}^{2}}{\sigma_{Y}^{2} + \sigma_{\hat{Y}}^{2} + (\mu_{Y} - \mu_{\hat{Y}})^2}, 
\end{equation}
where $Y$ and $\hat{Y}$ are the true and predicted labels respectively. $\mu_{Y}$, $\mu_{\hat{Y}}$, $\sigma_{Y}$, $\sigma_{\hat{Y}}$ and $\sigma_{Y\hat{Y}}$ refer to their means, variances and covariance respectively.  CCC $\in$ [-1,1], where -1 indicates total discordance (perfect negative relationship between predicted and true labels), 0 no concordance (no relationship between predicted and true labels) and +1 the perfect concordance (perfect positive relationship between predicted and true labels).

%%%%%%%%%%%%%%%%%%%%%%%%%%%%%%%%%%%
\subsection{Datasets}

CMU-MOSI contains 2199 utterances with a sentiment label from 93 YouTube movie review videos in English language \cite{zadeh2016multimodal}. There are 89 distinct speakers (41 females, 48 males).  The training, validation and test sets have 52 (1151), 10 (296) and 31 (752) videos (utterances), respectively. The videos 
are segmented into utterances with each utterance's sentiment label scored between +3 (strong positive) to -3 (strong
negative) by 5 annotators. The average of these five annotations is taken as the sentiment polarity
to create two classes (positive and negative)~\cite{poria2017context, poria2017multi, ghosal2018contextual}. We use utterance level features produced by a convolutional neural network \cite{karpathy2014large}, 3D convolutional neural network \cite{ji20123d} and openSMILE \cite{eyben2010opensmile} for textual, visual and acoustic modalities, respectively~\cite{ghosal2018contextual}. The dimensions of these features are 100, 100 and 73 for textual, visual and acoustic, respectively.

RECOLA (used in the 2015 \cite{ringeval2015av+}, 2016 \cite{valstar2016avec} and 2018 \cite{ringeval2018avec} Audio/Visual Emotion Challenges)  contains  audio-visual  recordings  of  spontaneous  interactions  of  27  French-speaking  participants   in a  remote  collaborative  task.  The dataset is equally divided into three disjoint parts (training, development and testing) by balancing the gender, age and mother-tongue of the participants: each  part  consists  of  nine  unique  recordings split in 40 ms segments. Continuous arousal and valence annotations in the range [-1,1] are provided for each segment for the first 5 minutes of each recording. The annotations are averaged over six annotators, accounting for inter-evaluator agreement.  Since the test labels are not publicly available, we report the results on the development set. For a fair comparison with previous literature~\cite{valstar2016avec, han2017reconstruction, han2017hard, zhang2018dynamic, han2019emobed, rajan2021robust}, we have employed the same acoustic and visual features as the AVEC baselines: 88-D extended Geneva Minimalistic Acoustic ParameterSet (eGeMAPS) features extracted using openSMILE \cite{eyben2010opensmile}, LGBP-TOP based  168-D  visual-appearance  features  and  49 facial landmarks based  632-D  visual-geometric  features. For each feature type, one feature vector is extracted for each 40 ms segment. Separate features are provided for arousal and valence estimation tasks. Similar to \cite{chen2020transformer}, to ensure that clips $V$ are long enough to capture the contextual information and to increase the number of training samples, we split the 9 recordings in training set by applying a sliding window of 3s with hop-size 1s. Thus, for each recording in training set, 299 clips are used, each including $N=75$ segments $V_n$. For the development set, we use non-overlapping 3s sequences. Note that when continuously assessing emotions, the annotators  need  time  to  perceive  audio-visual  events  and  report  the  emotional  states \cite{mariooryad2014correcting}, thus all ground-truth labels were shifted backwards by 2.8s, with an assumption that the delay is invariant with annotators, annotator states, modalities, and tasks~\cite{han2019emobed,valstar2016avec}.

\subsection{Implementation Details}
All our models are developed using PyTorch \cite{NEURIPS2019_9015}. The network architecture for all models is kept generic except for a few modality-specific and dataset-specific differences. 

The front-end of the encoder consists of 2 Bi-GRU layers, except for the audio models in RECOLA and the text model in CMU-MOSI where only a single layer Bi-GRU was used. The number of neurons in each Bi-GRU layer is equal to the input feature dimensionality. Dropout rates (\{0.2,0.3,0.4,0.5,0.6,0.7\}) are optimized for each model depending on the characteristics of the input features. The number of neurons in the dense layer succeeding the Bi-GRU layers is 100 which is the dimension of the latent representations. This is because the transformer encoder layers do not change the feature dimensionality of their inputs~\cite{vaswani2017attention}. Two transformer encoder layers are then employed with 2 self-attention operations in each transformer encoder layer. The first and second dense layers after the MHA modules in each transformer encoder layer have 400 and 100 neurons, respectively. Dropout regularisation is applied to the residual connections, first dense layer after MHA module as well as the attention weights in the transformer encoder to prevent overfitting.

For the decoder, we use a single layer Bi-GRU for the RECOLA valence audio model, while for all other models, we use 2 layers of Bi-GRU. For RECOLA, the number of neurons in the decoder Bi-GRU layers is kept 250 for arousal models and 500 for valence audio models. This is because for the latter, the decoder has to map the 100 dimensional latent features to higher dimensional video features (632 and 168). Finally, the number of neurons in the decoder dense layer is kept equal to the feature dimensionality of the stronger modality it is mapping into.

The classifier module contains a ReLU activated input dense layer with 300 neurons and an output dense layer with a single neuron activated using Sigmoid or Tanh for classification or regression respectively. A dropout of rate 0.3 is applied after its first dense layer to prevent overfitting.

We use a batch size of 32 and the Adam \cite{kingma2014adam} optimizer for training all the uni-modal baselines and the CM-StEW models. We use a learning rate of 1e-4 and 1e-5 for experiments on CMU-MOSI and RECOLA, respectively. We keep $r_1 = r_2 = 0.001$ in eq. \ref{eq:8}-\ref{eq:9} which is within the recommended range of [1e-8,10]~\cite{andrew2013deep}.

\begin{table}[t]\centering
\ra{1.3}
\caption{Uni-modal baseline results on CMU-MOSI for binary sentiment classification task. KEY - \textit{Acc.}: binary classification accuracy, $\overline{F}_{1}$: weighted $F_{1}$ score}
\begin{tabular}{@{}llccccc@{}}\toprule
& Method & \textit{Acc.} & $\overline{F}_{1}$ \\
\midrule
\multirow{2}{*}[-3pt]{Textual} & Classifier & 74.9 & 75.2 \\ 
& Encoder + Classifier & 80.3 & 80.1 \\
\midrule
\multirow{2}{*}[-3pt]{Acoustic} & Classifier & 54.7 & 55.2  \\
& Encoder + Classifier & 61.0 & 60.0  \\
\midrule
\multirow{2}{*}[-3pt]{Visual} & Classifier & 51.5 & 52.2  \\
& Encoder + Classifier & 60.8 & 59.7  \\
\bottomrule
\end{tabular}
\label{tab:uni-cmu}
\end{table}

\subsection{Uni-modal Baselines}
For sentiment classification, we assess the uni-modal performance of each modality with a classifier only model (the 2 dense layer model $\Gamma_s^c$ in Fig.\ref{fig:Model}) and an encoder with classifier model ($\Phi_s^e$ - $\Gamma_s^c$ in Fig.\ref{fig:Model}). Results are reported in Tab.~\ref{tab:uni-cmu}.
In accordance with previous works which found linguistic features to be more discriminative~\cite{seo2020hmtl,ghosal2018contextual,poria2017context}, we confirmed that the uni-modal performance of textual features surpasses that of acoustic or visual. Thus, we consider two cases for stronger-to-weaker  cross-modal knowledge transfer, namely, textual to acoustic and textual to visual.

Similar to sentiment classification, for emotion regression, we assess the uni-modal performance of acoustic, visual-geometric and visual-appearance features with a classifier only model (the 2 dense layer model $\Gamma_s^c$ in Fig.\ref{fig:Model}) and an encoder with classifier model ($\Phi_s^e$ - $\Gamma_s^c$ in Fig.\ref{fig:Model}). Results are reported in Tab.~\ref{tab:uni-recola}.
For arousal,  acoustic modality is  stronger  than  visual modality whereas for valence, visual modality  performs better than  acoustic modality.  These  results  are consistent  with  previous  studies  which  found  that  acoustic  and visual  features  are  more  discriminative  for  arousal  and  valence respectively~\cite{valstar2016avec,han2019emobed}. Hence, for stronger-to-weaker cross-modal knowledge transfer,  we  consider acoustic to visual-appearance and acoustic to visual-geometric for arousal  and  visual-appearance to acoustic  and  visual-geometric to acoustic for  valence,  respectively.

The results obtained using the two uni-modal baseline models in Tab.~\ref{tab:uni-cmu} and Tab.~\ref{tab:uni-recola} for both sentiment classification and emotion regression clearly indicate the performance improvement obtained by the addition of the encoder block to the classifier module. This validates our hypothesis that incorporating contextual information using the recurrence and attention mechanisms from the Bi-GRU and transformer encoder can help in better understanding the underlying affective behaviour.

\begin{table}[t]\centering
\ra{1.3}
\caption{Uni-modal baseline results in terms of Concordance Correlation Coefficient (CCC) on RECOLA for continuous emotion regression task. Note that the range of CCC is [-1,1].}
\begin{tabular}{@{}llccc@{}}\toprule
& Method & Acoustic & Visual & Visual \\ 
& & eGeMAPS & appearance & geometric \\
\midrule
\multirow{2}{*}[-3pt]{Arousal} & Classifier & 0.769 & 0.517 & 0.470 \\
& Encoder + Classifier & 0.786 & 0.541 & 0.536 \\
\midrule
\multirow{2}{*}[-3pt]{Valence} & Classifier & 0.490 & 0.496 & 0.560 \\
& Encoder + Classifier & 0.525 & 0.570 & 0.601 \\
\bottomrule
\end{tabular}
\label{tab:uni-recola}
\end{table}

\subsection{Cross-modal knowledge transfer}

In addition to our uni-modal baselines, we compare the performance of CM-StEW on CMU-MOSI using four uni-modal models (CAT-LSTM-Uni~\cite{poria2017context}, MU-SA~\cite{ghosal2018contextual}, Seq2SeqSentiment-Uni~\cite{pham2018seq2seq2sentiment}, HMTL-Uni~\cite{seo2020hmtl}) and two cross-modal knowledge transfer frameworks (Seq2SeqSentiment~\cite{pham2018seq2seq2sentiment} and HMTL~\cite{seo2020hmtl}). 
Since \cite{poria2017context} and \cite{ghosal2018contextual} did not report \textit{Acc.} and $\overline{F}_{1}$ scores respectively, we used their publicly available codes to asses the performance. For experiments on RECOLA, we compare with five uni-modal models (SVR+offset~\cite{valstar2016avec}, MTL(RE)~\cite{han2017reconstruction}, MTL(PU)~\cite{han2017hard}, DDAT(RE)~\cite{zhang2018dynamic} and DDAT(PU)~\cite{zhang2018dynamic}) and two cross-modal knowledge transfer frameworks (EmoBed~\cite{han2019emobed}, SEW~\cite{rajan2021robust}). For both  datasets, the methods chosen for the comparison have the best reported results on the weaker modalities. We considered only methods that reported results on the same set of features and the same dataset partition as we used.

For the sentiment classification task, Tab. \ref{tab:sota-cmu} reports the results obtained using CM-StEW when the weaker acoustic and visual modalities are improved using textual modality during training. First of all, we observe that our uni-modal baseline model, comprising the encoder and classifier, performs comparable to or better than four uni-modal models from the literature. This makes the improvement provided by CM-StEW extremely significant as the uni-modal models are very competitive.
Compared to the best uni-modal baseline results, CM-StEW improved the performance of both acoustic and visual models in terms of both accuracy (by 2.6\% and 3.8\% for visual and acoustic respectively) and $\overline{F}_1$ (by 4.2\% and 4.3\% for visual and acoustic respectively) metrics. Also considering the cross-modal knowledge transfer frameworks HMTL~\cite{seo2020hmtl} and Seq2SeqSentiment(+Textual)~\cite{pham2018seq2seq2sentiment}, except for uni-modal visual model accuracy, our models achieve better results, thus validating the effectiveness of knowledge transfer from the richer textual modality via the proposed methodology. Finally, the ablation results are obtained by removing the cross-modal decoder (- Decoder) or by removing the latent feature alignment mechanism (- LFA). It can be seen that both the decoder as well as LFA contribute towards the CM-StEW knowledge transfer process, with the contribution of LFA component being slightly higher than the decoder component.  

\begin{table}[t]\centering
\ra{1.3}
\caption{Results obtained using our proposed CM-StEW on CMU-MOSI in terms of Binary Accuracy (\textit{Acc.}) and Weighted $F_{1}$ ($\overline{F}_{1}$).* indicates results obtained using our evaluation on the publicly available codes.}% KEY - LFA: Latent Feature Alignment, Uni: uni-modal versions of the model, Sent.: sentiment, CAT: Contextual ATtention, MU-SA: Multi-Utterance Self Attention, HMTL: Heterogeneous Modality Transfer Learning.}
\begin{tabular}{@{}llcccc@{}}\toprule
Ref. & Method & \multicolumn{2}{c}{Visual} & \multicolumn{2}{c}{Acoustic} \\
%\addlinespace[0pt]
\cmidrule(lr){3-4} \cmidrule(lr){5-6}
%\addlinespace[-10pt]
& & \textit{Acc.} & $\overline{F_{1}}$ & \textit{Acc.} & $\overline{F_{1}}$ \\
%\midrule
%Ours & Best uni-modal & 60.8 & 59.7 & 61.0 & 60.0 \\
\midrule
%\multicolumn{6}{l}{\em{Other knowledge transfer methods}} \\ 
\cite{poria2017multi} & CAT-LSTM-Uni & - & 55.5 & - & 60.1 \\
\cite{poria2017multi} & CAT-LSTM-Uni* & 55.0 & 55.6 & 62.1 & 60.2 \\
\cite{ghosal2018contextual} & MU-SA & 63.7 & - & 62.1 & -\\
\cite{ghosal2018contextual} & MU-SA* & 62.8 & 61.9 & 59.7 & 58.4\\
\cite{pham2018seq2seq2sentiment} & Seq2SeqSent.-Uni & - & 48.0 & - & 56.0\\
\cite{seo2020hmtl} & HMTL-Uni & 62.1 & 61.3 & 58.2 & 58.2 \\
\hline
- & our uni-modal (Tab.~\ref{tab:uni-cmu}) & 60.8 & 59.7 & 61.0 & 60.0\\
\hline
\cite{pham2018seq2seq2sentiment} & Seq2SeqSent.(+Textual) & - & 58.0 & - & 56.0\\
\cite{seo2020hmtl} & HMTL(+Textual) & \textbf{64.8} & 61.7 & 62.6 & 60.8\\
\midrule
& CM-StEW(+Textual) & 63.4 & \textbf{63.9} & \textbf{64.8} & \textbf{64.3} \\
& \hspace{4mm} - LFA & 62.8 & 63.3 & 63.6 & 63.2 \\
& \hspace{4mm} - Decoder & 63.0 & 63.4 & 64.1 & 63.5 \\
\bottomrule
\end{tabular}
\label{tab:sota-cmu}
\end{table}

For the emotion regression task, Tab.~\ref{tab:sew-arousal} shows the results obtained on both arousal and valence estimation. For arousal, we can see that, with respect to other uni-modal methods from the literature, our uni-modal model provides better or comparable performance for both types of video features. This might be attributed to the fact that, unlike the compared models, our model takes contextual information into account. CM-StEW further improves the performance of the uni-modal baselines for both types of visual modality models. Specifically, an improvement of 0.033 and 0.050 are obtained for visual-appearance and visual-geometric models respectively, which are 6.1\% and 9.3\% of improvement over their best uni-modal baselines. Comparing our improved visual modality models with their counterparts obtained using other cross-modal knowledge transfer frameworks (EmoBed~\cite{han2019emobed}, SEW~\cite{rajan2021robust}), we observe that our models perform better than both EmoBed and SEW, thus validating the effectiveness of our knowledge transfer method. Finally, the ablation results indicate that both decoder as well as LFA contribute towards the CM-StEW method, with the contribution of decoder being slightly higher than the LFA component.

\begin{table}\centering
\ra{1.3}
\caption{Results obtained using our proposed CM-StEW on RECOLA  for Arousal predictions in terms of Concordance Correlation Coefficient (CCC).}% KEY - LFA: Latent Feature Alignment, MTL: Multi-Task Learning, RE: Reconstruction Error, PU: Perceptual Uncertainty, DDAT: Dynamic Difficulty Awareness Training, SVR: Support Vector Regressor, SEW: Stronger Enhancing Weaker, app: appearance, geo: geometric}
\begin{tabular}{@{}llccc@{}}\toprule
Ref. & Method & \multicolumn{2}{c}{Arousal} & Valence\\
\cmidrule(lr){3-4} \cmidrule(lr){5-5}
& & Visual & Visual & Acoustic \\
& & appearance & geometric & eGeMAPS \\
\midrule
\cite{valstar2016avec} & SVR + offset & 0.379 & 0.483 & 0.455 \\
\cite{han2017reconstruction} & MTL (RE) & 0.502 & 0.512 & 0.519 \\
\cite{han2017hard} & MTL (PU) & 0.508 & 0.502 & 0.506 \\
\cite{zhang2018dynamic} & DDAT (RE) & 0.544 & 0.539 & 0.508 \\
\cite{zhang2018dynamic} & DDAT (PU) & 0.513 & 0.518 & 0.498 \\
\hline
- & our uni-modal (Tab.~\ref{tab:uni-recola}) & 0.541 & 0.536 & 0.525 \\
\hline
\cite{han2019emobed} & EmoBed(+Acoustic) & 0.527 & 0.549 & - \\
\cite{han2019emobed} & EmoBed(+Visual-app.) & - & - & 0.514 \\
\cite{han2019emobed} & EmoBed(+Visual-geo.) & - & - & 0.521 \\
\cite{rajan2021robust} & SEW(+Acoustic) & 0.565 & 0.544 & - \\
\cite{rajan2021robust} & SEW(+Visual-geo.) & - & - & 0.552 \\
\midrule
%\multirow{3}{*}[-3pt]{Ours} 
& CM-StEW(+Acoustic) & \textbf{0.574} & \textbf{0.586} & - \\
& \hspace{4mm} - LFA & 0.571 & 0.580 & - \\
& \hspace{4mm} - Decoder & 0.561 & 0.564 & - \\
%\cmidrule{2-5}
%& $\Delta$ Best uni-modal & +0.033 & +0.050 & - \\
\midrule
%\multirow{3}{*}[-3pt]{Ours} 
& CM-StEW(+Visual-app.) & - & - & 0.560 \\
& \hspace{4mm} - LFA & - & - & 0.538 \\
& \hspace{4mm} - Decoder & - & - & \textbf{0.563} \\
%\cmidrule{2-5}
%& $\Delta$ Best uni-modal & - & - & +0.035 \\
\midrule
%\multirow{3}{*}[-3pt]{Ours} 
& CM-StEW(+Visual-geo.) & - & - & 0.555 \\
& \hspace{4mm} - LFA & - & - & 0.531 \\
& \hspace{4mm} - Decoder & - & - & \textbf{0.556} \\
%\cmidrule{2-5}
%& $\Delta$ Best uni-modal & - & - & +0.030 \\
\bottomrule
\end{tabular}
\label{tab:sew-arousal}
\end{table}

For valence estimation, Tab.~\ref{tab:sew-arousal} shows that the performance of the weaker acoustic modality model could be improved using a visual-appearance modality model or  a visual-geometric modality model. Specifically, we obtain an improvement of 0.035 and 0.030 when performing knowledge transfer from video-appearance and video-geometric models respectively, which are improvement  of 6.7\% and 5.7\% over the uni-modal acoustic baseline model (which in turn outperforms other uni-modal methods in the literature).  Comparison with the corresponding models from SEW and EmoBed, shows that our CM-StEW models achieve the best results in terms of CCC. Our ablation study shows that even though addition of both decoder and LFA components to the uni-modal baseline model improve performance, the absence of the decoder results in the best performing model for both cases. We hypothesise that this might be due to the presence of zero frames (features corresponding to frames where the face detector failed). The decoder might be mapping multiple acoustic features to the same visual features (zeros) thus decreasing the discriminative ability of the intermediate features. Nevertheless, the results are comparable to those obtained with the full CM-StEW model.

%%%%%%%%%%%%%%%%%%%%%%%%
\section{conclusion} \label{sec-6}

We presented a novel cross-modal knowledge transfer framework based on cross-modal translation and correlation based latent feature alignment. The proposed strategy used representations of a stronger modality as a guide for a weaker modality model to learn more discriminative representations. To account for contextual information, we used Bi-GRU and transformer encoder architecture as backbone. Experiments on two tasks (binary sentiment classification and continuous emotion regression) showed the effectiveness of our method that achieves new state-of-the-art results for the uni-modal testing with weaker modalities.

As future work, we will extend CM-StEW by making it agnostic to modality ranking and enable cross-modal knowledge transfer for test-time performance improvement by training using all available modalities. 

\ifCLASSOPTIONcaptionsoff
  \newpage
\fi



%\bibliographystyle{IEEEtran}
%\begin{thebibliography}{1}


%\end{thebibliography}

\bibliographystyle{IEEEbib}
\bibliography{main.bbl}

%\begin{IEEEbiography}{Vandana Rajan}
%\end{IEEEbiography}

%\begin{IEEEbiographynophoto}{Alessio Brutti}

%\newpage

%\begin{IEEEbiographynophoto}{Andrea Cavallaro}


%\vfill


\end{document}